\documentclass[twocolumn,prd,floatfix,showpacs]{revtex4}
\usepackage{graphicx}
\usepackage{epsfig}

\begin{document}

\title{Constraints on the sound speed of dark energy}
\author{Steen Hannestad}
\affiliation{Department of Physics and Astronomy, University of Aarhus, DK-8000 Aarhus C, Denmark}
\email{sth@phys.au.dk}
\date{\today}

\begin{abstract}
We have studied constraints on the equation of state, $w$, and speed of sound, $c_s$, of the dark energy from a joint analysis of data from the cosmic microwave background, large scale structure and type-Ia supernovae.
We find that current observations have no significant sensitivity to $c_s$. However, there is a slight difference between models in which there are no dark energy perturbations and models in which dark energy behaves as a fluid.
Assuming that there are no dark energy perturbations shifts the allowed region for $w$ to slightly higher values. 
At present models with and without dark energy perturbations provide roughly equally good fits to observations, but the difference is potentially important for future parameter estimations.
Finally, we have also performed error forecasts for future measurements of $c_s$.
\end{abstract}

\maketitle

\section{Introduction} 

The discovery in 1998 from observations of type Ia supernovae \cite{Riess:1998cb,Perlmutter:1998np} that the universal expansion is currently accelerating was a spectacular result. The finding has since been confirmed by observations of the Cosmic Microwave Background (CMB) \cite{Bennett:2003bz,Spergel:2003cb} and the large scale structure (LSS) of the universe \cite{2dFGRS,Tegmark:2003uf,Tegmark:2003ud}

One possible explanation is that the energy density of the universe is dominated by dark energy with a negative equation of state.
The simplest possibility is the cosmological constant which has $P=w\rho$ with $w=-1$ at all times. However, since the cosmological constant has a magnitude completely different from theoretical expectations one is naturally led to consider other explanations for the dark energy.

A light scalar field rolling in a very flat potential would for instance have a strongly negative equation of state, and would in the limit of a completely flat potential lead to $w=-1$ \cite{Wetterich:1987fm,Peebles:1987ek,Ratra:1987rm}. Such models are generically known as quintessence models. The scalar field is usually assumed to be minimally coupled to matter, but very interesting effects can occur if this assumption is relaxed (see for instance \cite{Mota:2004pa}).

In general such models would also require fine tuning in order to achieve $\Omega_X \sim \Omega_m$, where $\Omega_X$ and $\Omega_m$ are the dark energy and matter densities at present.
However, by coupling quintessence to matter and radiation it is possible to achieve a tracking behavior of the scalar field so that $\Omega_X \sim \Omega_m$ comes out naturally of the evolution equation for the scalar field \cite{Zlatev:1998tr,Wang:1999fa,Steinhardt:1999nw,Perrotta:1999am,%
Amendola:1999er,Barreiro:1999zs,Bertolami:1999dp,Baccigalupi:2001aa,%
Caldwell:2003vp}.

Many other possibilities have been considered, like $k$-essence, which is essentially a scalar field with a non-standard kinetic term \cite{Armendariz-Picon:1999rj,Chiba:1999ka,Armendariz-Picon:2000ah,%
Chimento:2003ta,Gonzalez-Diaz:2003rf,Scherrer:2004au,Aguirregabiria:2004te}.
It is also possible, although not without problems, to construct models which have $w<-1$, the so-called phantom energy models 
\cite{Caldwell:1999ew,Schulz:2001yx,Carroll:2003st,Gibbons:2003yj,%
Caldwell:2003vq,Nojiri:2003vn,Singh:2003vx,Dabrowski:2003jm,Hao:2003th,%
Stefancic:2003rc,Cline:2003gs,Brown:2004cs,Onemli:2002hr,Onemli:2004mb,%
Vikman:2004dc,polarski}.

Finally, there are even more exotic models where the cosmological acceleration is not provided by dark energy, but rather by a modification of the Friedman equation due to modifications of gravity on large scales \cite{Deffayet:2001pu,Dvali:2003rk}.

If dark energy is a fluid, i.e.\ a component with energy density and pressure, it can in general be characterized by one parameter in addition to the equation of state, $w \equiv P/\rho$, namely the speed of sound, $c_s$. Scalar field models with standard kinetic terms all have $c_s=1$ which is a solid prediction. On the other hand, models such as $k$-essence can have an effective sound speed which at times is $c_s \gg 1$ (note that this in fact does not violate causality). Models where the acceleration is due to a modification of gravity have no definable sound speed, and the "dark energy" component does not have perturbations.
Measuring the effective sound speed of dark energy would therefore provide crucial information which is complementary to the measurement of the equation of state. Note that another possibility is to discuss the dark energy parameters purely in terms of observables instead. This approach was for instance adobted in the recent paper \cite{Knox:2005rg}.

In the present paper we discuss current constraints on the sound speed of dark energy as well as the prospects for a future detection of this parameter. In Section II we review the formalism needed to describe perturbation evolution of a general fluid with equation of state and sound speed. In Section III we review the main observational probes of dark energy, and in section IV we discuss the current constraints on $w$ 
and $c_s$. In Section V we discuss possible future constraints on the sound speed from observations of the CMB, large scale structure, and weak lensing. Finally, section VI contains a discussion of the results.

\section{Formalism} 

The standard treatment of cosmological perturbation theory is the now classical paper by Ma and Bertschinger \cite{mb}, which described perturbation equations for general fluids in both synchronous and conformal gauge. The most commonly used numerical Boltzmann solver is CMBFAST \cite{CMBFAST} which uses synchronous gauge throughout. For that reason we follow the notation of \cite{mb}, but only derive the perturbation equations in synchronous gauge because we use a modified version of CMBFAST for all numerical computations.
In this gauge the perturbation equations for a general, imperfect fluid
have been derived several times in the literature (see for instance \cite{Hu:1998kj,Hu:1998tk,Erickson:2001bq,Bean:2003fb}). Here we simply review the equations as derived in Ref.~\cite{Bean:2003fb}.

The adiabatic speed of sound for any fluid is given by
\begin{equation}
c_{a}^2 = \frac{\dot{P}}{\dot{\rho}} = w - \frac{\dot{w}}{3H(1+w)}
\end{equation}
However, for an imperfect fluid the speed of sound is given by the more general relation
\begin{equation}
c_s^2 = \frac{\delta P}{\delta \rho}.
\end{equation}
If $c_s \neq c_a$, entropy perturbations will develop, determined by 
\begin{equation}
w \Gamma = \frac{\dot{p}}{\rho} \left(\frac{\delta p}{p} - \frac{\delta \rho}{\rho}\right).
\end{equation}
While many components in themselves behave as perfect fluids (dust, radiation, etc), the combination of two perfect fluids with different speeds of sound can produce an imperfect fluid in which $c_s \neq c_a$.

Many models of dark energy can also be described as imperfect fluids. For example a single scalar field model always has $c_s = 1$, while $c_a$ can be very different.

In synchronous gauge the density and velocity perturbations take a particularly simple form \cite{Bean:2003fb} 

\begin{eqnarray}
\dot{\delta} & = & -(1+w) \left[(k^2 + 9H^2(c_s^2-c_a^2) \frac{\theta}{k^2} + \frac{\dot h}{2}\right], \nonumber \\
\dot \theta & = & -H(1-3 c_s^2) \theta + \frac{c_s^2}{1+w} \delta - k^2 \sigma, \label{eq:pert}
\end{eqnarray}
where $\sigma$ is the anisotropic stress. This term arises from the next order in the Boltzmann hierarchy. In most cases this term can be neglected, and the Boltzmann equation can be truncated to form a closed system of equations, similar to how the Enskog expansion is performed \cite{Hannestad:2000gt}.
These equations are then straightforward to solve, provided that we know $w$ and $c_s^2$. It should be noted that the above equations are valid even for time-varying $w$.

However, our approach will now be to take $w$ to be constant, as has been done in many previous analyses \cite{Corasaniti:2001mf,Bean:2001xy,Hannestad:2002ur,Melchiorri:2002ux}.  At present there is no indication that $w$ is varying. Even though the present Type Ia supernova data seem to favour a rapid evolution of $w$, this indication vanishes if all available cosmological data is analysed \cite{wang04,hm,upadhye} (for other discussions of a time-varying $w$, see for instance \cite{Corasaniti:2004,Gong:2004a,Gong:2004b,Nesseris:2004wj,Feng:2004ad,Alam:2004a,Alam:2004b,Corasaniti:2002,Jassal:2004ej,Choudhury:2003tj,Huterer:2004,Daly:2003,Wang:2004a,Wang:2004b,Wang:2004c,Jonsson:2004,Weller:2003hw,Linder:2003}.

$c_s^2$ is also taken to be a constant in our analysis. Although it is possible to produce models with extremely rapidly varying speed of sound, such as some $k$-essence models, as it will turn out the present constraints on $c_s^2$ are so weak that adding parameters to describe a time variation of $c_s^2$ is not justified at present.

Finally, it should be noted that the analysis can be carried to $w<-1$ without any problems, as long as $w$ is constant and $c_s^2 >0$. In that case there are no singularities in the perturbation equations and they can be solved without any problems. We therefore also include this region in the likelihood analysis in the next section.

From the perturbation equations it is clear that for $w=-1$ there can be no evolution of perturbations in the dark energy since $\dot \delta =0$ at all times (for the possibility of initial dark energy perturbations, see for instance \cite{Gordon:2004ez,Gordon:2005vz}).

Furthermore, if $w$ is sufficiently negative, dark energy only dominates at $z \sim 0$. This means that the speed of sound can only affect the very largest observable scales for these models.

\section{Observational probes} 

\subsection{The cosmic microwave background}

The temperature fluctuations are conveniently described in terms
of the spherical harmonics power spectrum $C_{T,l} \equiv \langle
|a_{lm}|^2 \rangle$, where $\frac{\Delta T}{T} (\theta,\phi) =
\sum_{lm} a_{lm}Y_{lm}(\theta,\phi)$.  Since Thomson scattering
polarizes light, there are also power spectra coming from the
polarization. The polarization can be divided into a curl-free
$(E)$ and a curl $(B)$ component, yielding four independent power
spectra: $C_{T,l}$, $C_{E,l}$, $C_{B,l}$, and the $T$-$E$
cross-correlation $C_{TE,l}$.

\begin{figure}[t]
\vspace*{-0.0cm}
\begin{center}
\epsfysize=9truecm\epsfbox{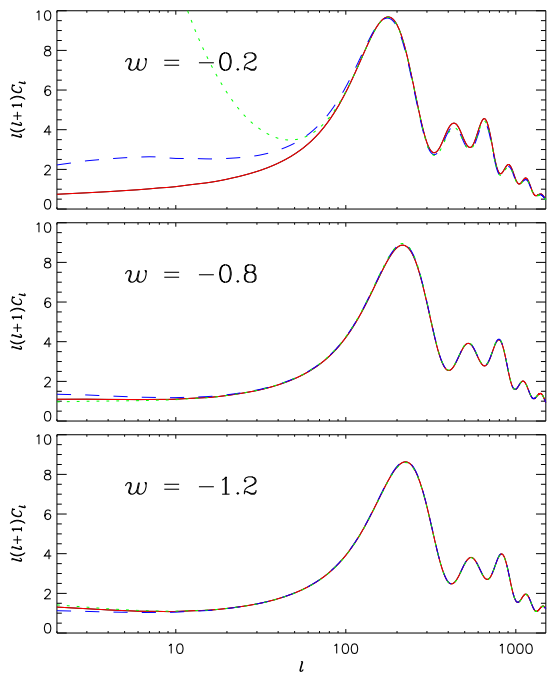}
\vspace{0.5truecm}
\end{center}
\caption{The CMB spectra for three different values of $w$. The full lines are for $c_s^2 = 0$, the dashed for $c_s^2 = 1$, and the dotted line is without dark energy perturbations.} \label{fig:cmb}
\end{figure}

The speed of sound of dark energy mainly affects the CMB spectrum at the very largest scales via the late Integrated Sachs Wolfe effect. In Fig.~\ref{fig:cmb} we show CMB spectra for several different models. From the figure it is evident that there can only be significant changes when $w$ is close to 0. It should also be noted that when there are no dark energy perturbations (as in the case where the acceleration is caused by modified gravity or other non-fluid like effects) the CMB spectrum in general is different. From the perturbation equations it is clear that there is no simple solution where $\dot \delta = 0$ at all times unless $w=-1$. The reason is that, independent of $c_s^2$, any fluid is affected by the metric term $\dot h$.
In principle it will therefore be possible to distinguish any model in which the dark matter behaves as a fluid from models where there are no dark energy perturbations. However as $w$ approaches -1 the difference vanishes.

\subsection{Large scale structure}

Any large scale structure survey measures the correlation function between galaxies. In the linear regime where fluctuations are Gaussian the fluctuations can be described by the galaxy-galaxy power spectrum alone, $P(k) = |\delta_{k,gg}|^2$. In general the the galaxy-galaxy power spectrum is related to the matter power spectrum via a bias parameter, $b^2 \equiv P_{gg}/P_m$. In the linear regime, the bias parameter is approximately constant, so up to a normalization constant $P_{gg}$ does measure the matter power spectrum.

\begin{figure}[t]
\vspace*{-0.0cm}
\begin{center}
\epsfysize=9truecm\epsfbox{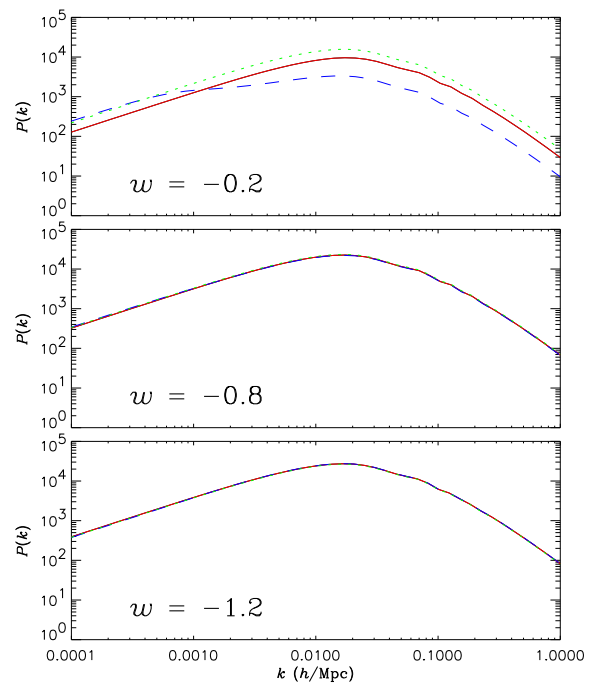}
\vspace{0.5truecm}
\end{center}
\caption{The matter spectra for three different values of $w$. The full lines are for $c_s^2 = 0$, the dashed for $c_s^2 = 1$, and the dotted line is without dark energy perturbations.} \label{fig:lss}
\end{figure}

In Fig.~\ref{fig:lss} we show matter power spectra for several different models. Again, the differences are negligible as soon as $w$ approaches $-1$.

\subsection{Weak lensing}

\begin{figure}[t]
\vspace*{-0.0cm}
\begin{center}
\epsfysize=9truecm\epsfbox{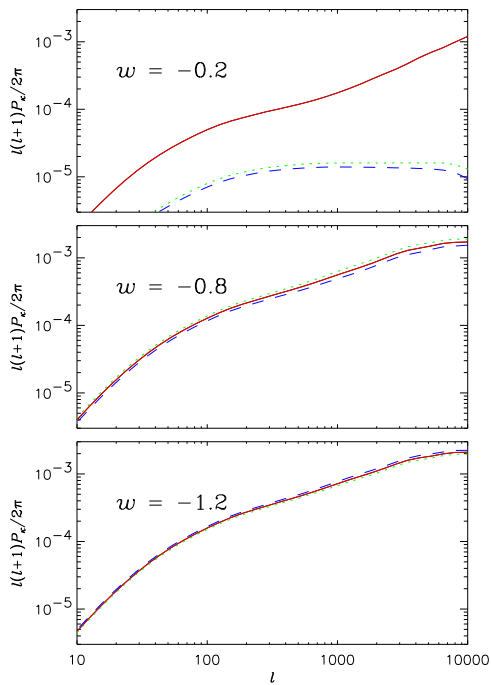}
\vspace{0.5truecm}
\end{center}
\caption{The lensing spectra for three different values of $w$. The full lines are for $c_s^2 = 0$, the dashed for $c_s^2 = 1$, and the dotted line is without dark energy perturbations.} \label{fig:lensing}
\end{figure}

At present there are no large scale weak lensing surveys. However, in the future weak lensing will be one of the main probes of cosmology. The advantage of measuring weak lensing is that it measures evolution at late times where dark energy contributes significantly to the energy density. 
In a flat universe the angular convergence power spectrum of weak lensing is, in the normally used Limber approximation, given by \cite{js}
\begin{equation}
P_\kappa = \frac{9}{4} H_0^4 \Omega_m^2 \int_0^{\chi_H} \frac{g^2(\chi)}{a^2(\chi} P_{\rm nl} \left(\frac{l}{\chi},\chi\right) d\chi.
\end{equation}
Here, $H_0$ is the present Hubble parameter, $\Omega_m$ is the matter density, $l$ is the multipole moment, and $\chi$ is the radial distance.
$P_{\rm nl}$ is the non-linear matter power spectrum at wave number $l/\chi$ and redshift corresponding to a horizon distance of $\chi$.
$g(\chi)$ is the lensing probability function
\begin{equation}
g(\chi) = \int_{\chi}^{\chi_H} n(\chi') \frac{\chi'-\chi}{\chi'}d\chi.
\end{equation}
$n(\chi)$ is the distribution function of sources, normalised to $\int n(z) dz=1$. Since sources exist at a distribution of redshifts which are measurable, it is even possible to divide them into redshift bins and do dark energy tomography.
The CMB itself can also be used as a source in which case $n$ can be taken as a delta function with $z \simeq 1100$.
We have calculated lensing spectra for the same models as before, assuming that all sources are located at $z = 2$. The non-linear power spectrum was calculated using the Peacock-Dodds formalism. This approach is semi-analytic and may not be accurate enough for real parameter extraction. However, for calculating differences between models it is appropriate, as discussed in Ref.~\cite{seljak}.
The results are shown in Fig.~\ref{fig:lensing}. As before the differences are significant only when $w$ is close to zero. However, as will be discussed in section V, future lensing surveys will improve the sensitivity to the sound speed of dark energy significantly.

\subsection{Type Ia supernovae}

Measurements of the cosmological relationship between luminosity distance and redshift in itself has ho sensitivity to the speed of sound of dark energy because it only probes the evolution of the homogeneous background. However, since these measurements significantly constrain other cosmological parameters they break some of the degeneracies between $c_s$ and other cosmological parameters.

\section{Current constraints} 

Using the presently available precision data we have performed a likelihood analysis for the two dark energy parameters $w$ and $c_s$.

As our framework we choose the
minimum standard model with 6
parameters: $\Omega_m$, the matter density, the curvature
parameter, $\Omega_b$, the baryon density, $H_0$, the Hubble
parameter, and $\tau$, the optical depth to reionization. The normalization of both CMB and LSS spectra are taken to be free and unrelated parameters.
The priors we use are given in Table~\ref{table:priors}.

\begin{table}
\begin{center}
\begin{tabular}{lcl}
\hline \hline
parameter & prior\cr
\hline
$\Omega=\Omega_m+\Omega_X$&1&Fixed\cr
$h$ & $0.72 \pm 0.08$&Gaussian \cite{freedman}\cr
$\Omega_b h^2$ & 0.014--0.040&Top hat\cr
$n_s$ & 0.6--1.4& Top hat\cr
$\tau$ & 0--1 &Top hat\cr
$Q$ & --- &Free\cr
$b$ & --- &Free\cr
\hline \hline
\end{tabular}
\end{center}
\caption{The different priors on parameters
used in the likelihood analysis.}
\label{table:priors}
\end{table}

Likelihoods are calculated from $\chi^2$ so that in 2-dimensional plots the 68\% and 95\% regions are formally
defined by $\Delta \chi^2 = 2.30$ and 6.17
respectively. Note that this means that the 68\% and 95\% contours
are not necessarily equivalent to the same confidence level for single
parameter estimates.

\subsection{Supernova luminosity distances}
We perform our likelihood analysis using the ``gold'' dataset compiled
and described in Riess et al \cite{Riess:2004} consisting of 157
SNIae using a modified version of the
SNOC package \cite{Goobar:2002c}.

\subsection{Large Scale Structure (LSS).}

At present there are two large galaxy surveys of comparable size, the
Sloan Digital Sky Survey (SDSS) \cite{Tegmark:2003uf,Tegmark:2003ud}
and the 2dFGRS (2~degree Field Galaxy Redshift Survey) \cite{2dFGRS}.
Once the SDSS is completed in 2005 it will be significantly larger and
more accurate than the 2dFGRS. In the present analysis we use data from SDSS, but the results would be almost identical had we used 2dF data instead. In the data analysis we use only data points on scales larger than $k = 0.15 h$/Mpc in order to avoid problems with non-linearity.

\subsection{Cosmic Microwave Background.}

The WMAP experiment has reported data on $C_{T,l}$ and $C_{TE,l}$
as described in
Refs.~\cite{Spergel:2003cb,Bennett:2003bz,Kogut:2003et,%
Hinshaw:2003ex,Verde:2003ey}.  We have performed our
likelihood analysis using the prescription given by the WMAP
collaboration~\cite{Spergel:2003cb,Bennett:2003bz,Kogut:2003et,%
Hinshaw:2003ex,Verde:2003ey} which includes the
correlation between different $C_l$'s. Foreground contamination has
already been subtracted from their published data.

\subsection{Results}

\begin{figure}[t]
\vspace*{-0.0cm}
\begin{center}
\epsfysize=7truecm\epsfbox{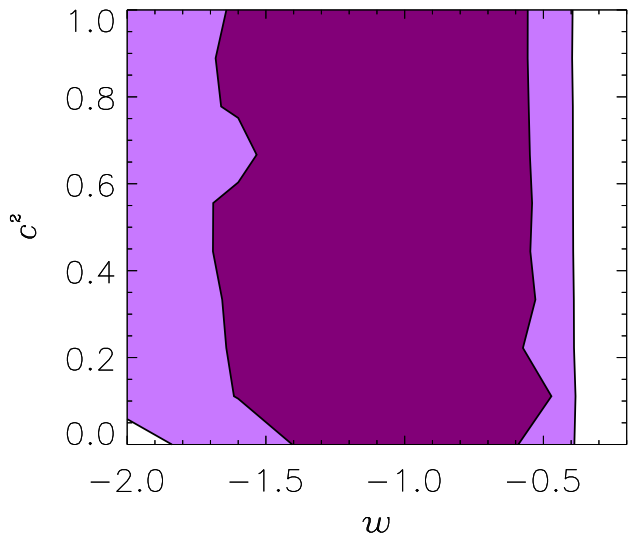}
\vspace{0.5truecm}
\end{center}
\caption{The 68\% (dark) and 95\% (light) likelihood contours for $w$ and $c_s^2$ for WMAP and SDSS data.} \label{fig4}
\end{figure}

In Fig.~\ref{fig4} we show the results of thew likelihood analysis for WMAP and SDSS data only. As is also found in many other analyses of the same data for $w$ alone, the data is compatible with $w=-1$. Furthermore there is no constraint on the speed of sound. Note here that the bound in $w$ comes from the combination of CMB and LSS data. If only CMB data is used then very high values of $w$ are allowed. This in turn means that as $w$ increases the sensitivity to $c_s^2$ also increases. This effect was found in the WMAP data by Bean and Dore \cite{Bean:2003fb}, and if only the WMAP data is considered there is a tentative indication that $c_s^2 < 1$. However, as soon as LSS data is added the allowed high $w$ region disappears and there are significant constraints only on $w$.

\begin{figure}[t]
\vspace*{-0.0cm}
\begin{center}
\epsfysize=7truecm\epsfbox{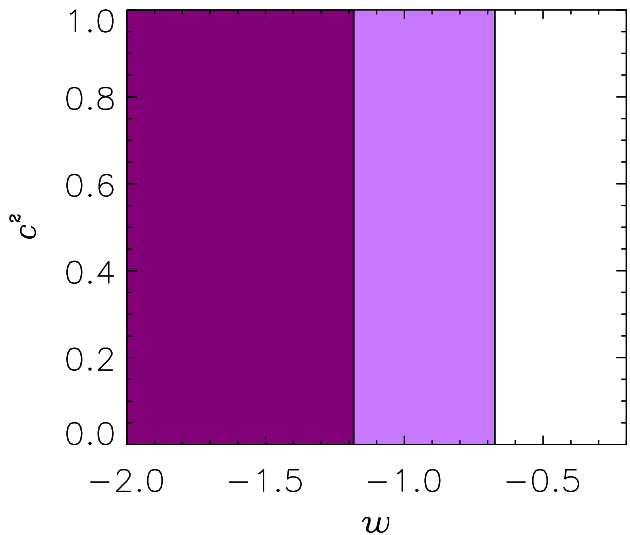}
\vspace{0.5truecm}
\end{center}
\caption{The 68\% (dark) and 95\% (light) likelihood contours for $w$ and $c_s^2$ for SNI-a data.} \label{fig5}
\end{figure}

Next, Fig.~\ref{fig5} shows the same analysis, but only using SNI-a data.
These measurements are only sensitive to $w$ so likelihood contours are vertical lines.

\begin{figure}[t]
\vspace*{-0.0cm}
\begin{center}
\epsfysize=7truecm\epsfbox{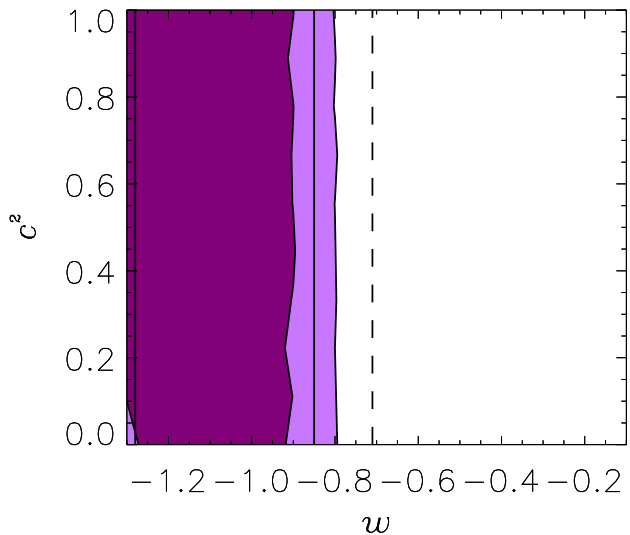}
\vspace{0.5truecm}
\end{center}
\caption{The 68\% (dark) and 95\% (light) likelihood contours for $w$ and $c_s^2$ for WMAP, SDSS, and SNI-a data. The vertical full lines are 68\% contours for the model with no dark energy perturbations, and the vertical dashed lines are 95\% contours for the same model.} \label{fig6}
\end{figure}

Finally, in Fig.~\ref{fig6} we show the likelihood analysis combining all current constraints. As can be seen there is again no sensitivity to $c_s^2$ because the constraint on $w$ is so strong that it rules out the region where changes in the speed of sound cause significant effects.

\subsubsection{Comparison with models with no dark energy perturbations}

However, it is also interesting to compare the likelihood analysis where the dark energy is a fluid with well-defined equation of state and speed of sound to one where there are no perturbations in the dark energy. As mentioned before there if no continuous transition between models with any finite $c_s$ and models with no dark energy perturbations because any fluid is subject to evolution in $\delta$ from the metric term $\dot h$ in the perturbation equations.
The difference between models with no dark energy perturbations and fluid models is therefore also quite interesting to probe.
The vertical lines in Fig.~\ref{fig6} show the likelihood analysis with the same data, but without dark energy perturbations. As can be seen the allowed region is shifted to slightly higher values of $w$. The best fit $\chi^2$ is 1626.1 for the model with no dark energy perturbations, whereas it is 1625.5 for the fluid model. Considering that the fluid model contains one more fitting parameter, there is no evidence at present for or against perturbations in the dark energy.
The model with no perturbations has 1516 degrees of freedom, whereas the fluid model has 1515. This means that $\chi^2/{\rm d.o.f} = 1.073$ in both cases.

\section{Future measurements} 

While there are no current constraints on the dark energy speed of sound, measurements will improve significantly in the future. We have performed a simple Fisher matrix error estimation of the performance of future measurements. The estimate is along the lines of that presented in Ref.~\cite{hu1}, but includes future large scale structure surveys.

It is possible to estimate the precision with which the cosmological
model parameters can be extracted from a given hypothetical data set.
The starting point for any parameter extraction is the vector of
data points, $x$. This can be in the form of the raw data, or in
compressed form, typically the power spectrum ($C_l$ for CMB and
$P(k)$ for LSS).

Each data point has contributions from both signal and noise,
$x = x_{\rm signal} + x_{\rm noise}$. If both signal and noise are
Gaussian distributed it is possible to build a likelihood function
from the measured data which has the following form \cite{oh}
\begin{equation}
{\cal L}(\Theta) \propto \exp \left( -\frac{1}{2} x^\dagger 
[C(\Theta)^{-1}] x \right),
\end{equation}
where $\Theta = (\Omega, \Omega_b, H_0, n_s, \tau, \ldots)$ is a vector
describing the given point in model parameter space and 
$C(\Theta) = \langle x x^T \rangle$ 
is the
data covariance matrix.
In the following we shall always work with data in the form of a
set of power spectrum coefficients, $x_i$, which can be either
$C_l$ or $P(k)$.

If the data points are uncorrelated so that the data covariance matrix
is diagonal, the likelihood function can be reduced to
${\cal L} \propto e^{-\chi^2/2}$, where
\begin{equation}
\chi^2 = \sum_{i=1}^{N_{\rm max}} \frac{(x_{i, {\rm obs}}-x_{i,{\rm theory}})^2}
{\sigma(x_i)^2},
\label{eq:chi2}
\end{equation} 
is a $\chi^2$-statistics and $N_{\rm max}$ 
is the number of power spectrum data
points \cite{oh}.

The maximum likelihood is an unbiased estimator, which means that
\begin{equation}
\langle \Theta \rangle = \Theta_0.
\end{equation}
Here $\Theta_0$ indicates the true parameter vector of the underlying
cosmological model and $\langle \Theta \rangle$ is the average estimate
of parameters from maximizing the likelihood function.

The likelihood function should thus peak at $\Theta \simeq \Theta_0$, and
we can expand it to second order around this value.
The first order derivatives are 
zero, and the expression is thus
\begin{widetext}
\begin{equation}
\chi^2  = \chi^2_{\rm min} + \sum_{i,j}(\theta_i-\theta) \left( \sum_{k=1}^{N_{\rm max}}
\frac{1}{\sigma (x_k)^2} \left[\frac{\partial x_k}{\partial \theta_i}
\frac{\partial x_k}{\partial \theta_j} - (x_{k, {\rm obs}}-x_k)
\frac{\partial^2 x_k}{\partial \theta_i \partial \theta_j} \right]\right) (\theta_j-\theta),
\end{equation}
\end{widetext}
where $i,j$ indicate elements in the parameter vector $\Theta$.
The second term in the second derivative can be expected to be very small
because $(x_{k, {\rm obs}}-x_k)$ is in essence just a random measurement error 
which should average out. The remaining term
is usually referred to as the Fisher information matrix
\begin{equation}
F_{ij} = \frac{\partial^2 \chi^2}{\partial \theta_i \partial \theta_j} = 
\sum_{k=1}^{N_{\rm max}}\frac{1}{\sigma (x_k)^2}\frac{\partial x_k}{\partial \theta_i}
\frac{\partial x_k}{\partial \theta_j}.
\label{eq:fisher1}
\end{equation}
The Fisher matrix is closely related to the precision with which the
parameters, $\theta_i$, can be determined.
If all free parameters are to be determined from the data alone without
any priors then it follows from the Cramer-Rao inequality
\cite{kendall} that
\begin{equation}
\sigma(\theta_i) = \sqrt{(F^{-1})_{ii}}
\end{equation}
for an optimal unbiased estimator, such as the maximum likelihood
\cite{tth}.

\subsection{Mock LSS surveys}

For purposes of parameter estimation the most important parameter
in galaxy surveys is the effective volume, defined as
\begin{equation}
V_{\rm eff} = \int \left[\frac{\bar{n}({\bf r}) P(k)}
{1+\bar{n}({\bf r}) P(k)}\right] d^3 r.
\end{equation}
In the above equation $n({\bf r})$ is the selection function. The
simple interpretation of $V$ is that it is the volume available
for measuring power at wavenumber $k$.

In the following we shall assume that the survey is volume limited,
meaning that the selection function is constant throughout the
survey volume. If the survey is flux limited the selection function
is much more complicated.
In the region where $P(k) \gtrsim 1/\bar{n}$, $V_{\rm eff}$ is
independent of $k$ and equal to the total survey volume.

Essentially this means that, with certain restrictions,
we can use just one free parameter,
$V_{\rm eff}$, to describe a hypothetical galaxy survey.

It was shown in Ref.~\cite{tegmark_lss} that the contribution
to the Fisher matrix from such a galaxy survey can be written as
\begin{equation}
F_{ij} \simeq 2 \pi \int_{k_{\rm min}}^{k_{\rm max}}
\frac{\partial \ln P(k)}{\partial \theta_i}
\frac{\partial \ln P(k)}{\partial \theta_j}
w(k) d \ln k,
\end{equation}
where the weight-function is $w(k) = V_{\rm eff}/\lambda^3$ and
$\lambda = 2\pi/k$. The upper limit of the integral should be
taken to be $k_{\rm cut}$, which is the scale where non-linearity starts to be significant.
In principle the lower limit, $k_{\rm min}$, should be zero 
but at large scales the assumption that 
$P(k) \gtrsim 1/\bar{n}$ breaks down. However, by far the most
of the weight in the above integral comes from $k$ close to the
upper limit. Therefore, as long as the $k$ where 
$P(k) = 1/\bar{n}$ is much smaller than $k_{\rm max}$ the error from
taking $V_{\rm eff}$ {\it and} $k_{\rm min} = 0$ is quite small.

It should be noted that the above integral expression is quite crude.
However, it offers a very simple way of estimating parameter estimation
errors from galaxy surveys. The error arising from it can be
of order a factor 2, 
leading to an error in the estimated $\sigma (\theta_i)$
of order $2^{1/2}$.

Instead of $V_{\rm eff}$ we use 
$\lambda_{\rm eff} = (3V_{\rm eff}/4\pi)^{1/3}$ as the free parameter.
As discussed in Ref.\ \cite{tegmark_lss} the SDSS Bright Red Galaxy (BRG) survey 
has an
effective volume of roughly $(1 h^{-1} \,\, {\rm Gpc})^3$, corresponding
to $\lambda_{\rm eff} \simeq 620 h^{-1} \,\, {\rm Mpc}$.

Note that the number of independent Fourier modes on a given scale, $k$,
enclosed within the survey volume
is proportional to $V_{\rm eff}$. 
Therefore it essentially corresponds to the factor $(2l+1)$ for the
CMB measurements which measures the number of $m$-modes for a given $l$.
In that sense both $V_{\rm eff}$ and $(2l+1)$ are a measure of the
lack of ergodicity in the given data set.

\subsection{Mock CMB experiments}

For a mock CMB experiment we assume it to be cosmic variance (as opposed to foreground) limited
up to some maximum $l$-value. This value can, however, be different
for temperature and polarization detection. Therefore a given hypothetical
experiment can be described by only two free parameters,
$l_{T, {\rm max}}$ and $l_{P, {\rm max}}$. For all experiments it will
be the case that $l_{T, {\rm max}} \geq l_{P, {\rm max}}$.

In this picture the MAP data will be well described by
$l_{T, {\rm max}} \simeq 1000$ and $l_{P, {\rm max}} = 0$, and
the Planck data by 
$l_{T, {\rm max}} \simeq 2500$ and $l_{P, {\rm max}} = 1500$.
In some sense Planck can therefore be regarded as the ``ultimate'' CMB
experiment because is measures all of the power spectrum parameter
space not dominated by foregrounds.

The contribution to the Fisher matrix from such a CMB experiment
is then

\begin{eqnarray}
F_{ij} & = & \sum_{l=2}^{l_{P,{\rm max}}}
\sum_{X,Y} \frac{\partial C_{l,X}}{\partial \theta_i}
{\rm Cov}^{-1} (C_{l,X},C_{l,Y})
\frac{\partial C_{l,Y}}{\partial \theta_j} \nonumber \\
&& \,\, + \sum_{l=l_{P,{\rm max}}}^{l_{T,{\rm max}}}
\frac{\partial C_{l,T}}{\partial \theta_i}
{\rm Cov}^{-1} (C_{l,T},C_{l,T})
\frac{\partial C_{l,T}}{\partial \theta_j},
\end{eqnarray}
where $X,Y = T,E,TE$.

The covariance matrices are given by \cite{wss}
\begin{eqnarray}
{\rm Cov}(C_{l,T},C_{l,T}) & = & \frac{2}{(2l+1)}C_{l,T}^2 \\
{\rm Cov}(C_{l,E},C_{l,E}) & = & \frac{2}{(2l+1)}C_{l,E}^2 \\
{\rm Cov}(C_{l,TE},C_{l,TE}) & = & \frac{2}{(2l+1)}[C_{l,T}^2
+ C_{l,T}C_{l,E}] \\
{\rm Cov}(C_{l,T},C_{l,E}) & = & \frac{2}{(2l+1)}C_{l,TE}^2 \\
{\rm Cov}(C_{l,T},C_{l,TE}) & = & \frac{2}{(2l+1)}C_{l,T}C_{l,TE} \\
{\rm Cov}(C_{l,E},C_{l,TE}) & = & \frac{2}{(2l+1)}C_{l,E}C_{l,TE}
\end{eqnarray}

It should be noted here that this approximation relies on
the assumption of $4\pi$ sky coverage and no pixel noise up to
the maximum $l$. Even though these assumptions are not realised in
any real experiment they are sufficiently accurate for estimating
the parameter estimation accuracy of a given experiment.

Finally, a word of caution on the treatment used here. In the Fisher
matrix analysis for CMB no window function is used, i.e.\ it is
assumed that all $C_l$ values are uncorrelated. While this is true
for $l$ much smaller than the resolution limit for a given experiment,
it breaks down close $l_{\rm max}$. Furthermore, since most of the
information on cosmological parameters, including the neutrino mass, 
comes from high $l$ this is a real concern.
However, by choosing $l_{\rm max}$ conservatively in the Fisher matrix
analysis this problem of overestimating precision at $l$ close to the
resolution limit can be minimized.

\subsection{Weak lensing measurements}

The uncertainty in the angular convergence spectrum for a weak lensing survey is given by
\begin{equation}
\sigma(P_\kappa) = \left(\frac{2}{(2l+1)f_{\rm sky}}\right)^{1/2} \left[P_\kappa + \frac{\langle \gamma_{\rm int}^2\rangle}{\bar{n}}\right],
\end{equation}
where $f_{\rm sky}$ is the fraction of sky measured. $\langle \gamma_{\rm int}^2\rangle \simeq 0.4$ is the intrinsic ellipticity of galaxies and $\bar{n} \simeq 6.6 \times 10^8 \,\, {\rm sr}^{-1}$ is the angular density of galaxies. At small scales non-linearity invalidates the assumption of Gaussianity inherent in the Fisher matrix formalism, and for this reason we do not use data beyond $l=3000$.
For a lensing survey of this type the only adjustable parameter is $f_{\rm sky}$ which can at maximum be about 0.65.

\subsection{Results}

In order to calculate estimated $1\sigma$ errors on the various
cosmological parameters we need to apply the Fisher matrix analysis
to a specific cosmological model.

We choose as the reference model the generic $\Lambda$CDM model with
the following free parameters:
$\Omega_m$, the matter density, $\Omega_b$, the baryon density, $H_0$, the Hubble parameter,
$n_s$, the spectral index of the primordial perturbation spectrum,
$\tau$, the optical depth to reionization, $Q$, the spectrum normalization,
$b$, the bias parameter, as well as $w$ and $c_s^2$.
The reference model has the following parameters:
$\Omega_m = 0.3$, $\Omega_{\rm DE} = 0.7$, $\Omega_b h^2 = 0.02$,
$H_0 = 70 \,\, {\rm km} \, {\rm s}^{-1} \, {\rm Mpc}^{-1}$,
$n_s = 1$, $\tau=0$, $w=-0.8$, and $c_s^2=1$.

As the mock data sets we take the following: For CMB we assume a cosmic variance limited experiment up to $l=2500$ for temperature and 1500 for polarization. For LSS we use two cases, one with 
$\lambda_{\rm eff} \simeq 620 h^{-1} \,\, {\rm Mpc}$ and one with $\lambda_{\rm eff} \simeq 2000 h^{-1} \,\, {\rm Mpc}$. In both cases we take $k_{\rm cut} = 0.1 \,h/$Mpc.
For weak lensing we take $f_{\rm sky} = 0.65$.

The estimated 1$\sigma$ error bars for the various cases are shown in Table \ref{table:fisher}. Even in the best case scenario $\sigma(c_s^2) \gg 1$ which means that there is little hope of measuring the actual value of the speed of sound.

However, it may still be possible to discern models with no dark energy perturbations from models with fluid-like dark energy.

\begin{table}
\begin{center}
\begin{tabular}{lc}
\hline \hline
Data & $\sigma(c_s^2)$ \cr
\hline
CMB only & 6.26 \cr
CMB + LSS$_{620}$ & 5.67\cr
CMB + LSS$_{2000}$ & 5.66 \cr
CMB + LSS$_{2000}$+WL & 3.03 \cr
\hline \hline
\end{tabular}
\end{center}
\caption{1$\sigma$ uncertainty on $c_s^2$ for different cosmological data sets.}
\label{table:fisher}
\end{table}

\section{Discussion} 

We have calculated constraints on the main dark energy parameters, the equation of state, $w$, and the speed of sound, $c_s^2$, from current observational data.

It was found that there is at present no useful constraint on the speed of sound of such a general fluid component. The reason is that current observations constrain $w$ to be close to $-1$ where the sound speed of dark energy has a negligible influence on perturbation evolution.

We also compared models in which dark energy behaves as a fluid to models with no dark energy perturbations. Here, it was found that the allowed region for $w$ is shifted to slightly higher values. The $\chi^2/{\rm d.o.f}$, however, is equal to 1.073 in both cases, meaning that at present there is no evidence for or against perturbations in the dark energy.

With regards to future observations we have performed a Fisher matrix analysis of CMB, LSS, and weak lensing measurements. Even in the most optimistic case where full sky measurements of CMB polarization and 
weak lensing can be performed the projected sensitivity to the speed of sound in fluid models is quite poor, and there is little hope of measuring it. However, it might be possible to discern between models where there are no dark energy perturbations and models in which the dark energy behaves as a fluid.

Finally we note that even if the speed of sound cannot be measured, it can still bias measurements of other parameters. For example, for the present cosmological data, treating the dark energy as a fluid leads to a slightly different estimate for the dark energy equation of state compared to the case where no dark energy perturbations are assumed.

\end{document}